\documentclass[12pt,preprint]{aastex}
\usepackage{epsfig,latexsym,longtable, graphicx}
\begin{document}


\def\sun{\hbox{$\odot$~}}
\def\deg{\hbox{$^\circ$}}
\def\kms{\,km\,s$^{-1}$}
\def\m{$^{\rm m}$}
\def\si{$\sim$}
\def\di{$\div$}
\def\av{A$_{\rm V}$ }
\def\msol{~M$_\odot$ }
\def\msolr{~M$_\odot$~yr$^{-1}$ }
\def\micron{\,$\mu$m}
\def\hi{H\,{\sc i} }
\def\marc{mag~arcsec$^{-2}$}
\def\cm2{cm$^{-2}$}
\def\ecs{ergs cm$^{-2}$ s$^{-1} \ $}
\def\es{ergs s$^{-1}$}
\def\cts{counts s$^{-1} \ $}
\def\esz{ergs s$^{-1}$ Hz$^{-1} \ $}
\def\kms{km s$^{-1}$}
\def\kcssk{$keV\ cm^{-2}\ s^{-1}\ sr^{-1}\ keV^{-1}\ $}
\def\ni {\noindent}
\def\Msun{$M_{\odot}\ $}
\def\lae{\mathrel{<\kern-1.0em\lower0.9ex\hbox{$\sim$}}}
\def\gae{\mathrel{>\kern-1.0em\lower0.9ex\hbox{$\sim$}}}

\title{The first optical  validation of an XLEO:\\ a detection in the XMM-{\it Newton} observation
of the CDFS\footnote{Based on observations obtained with XMM-{\it Newton}, an ESA science mission
with instruments and contributions directly funded by
ESA Member States and NASA.}}
\author{V. Braito\altaffilmark{1}, T. Maccacaro\altaffilmark{1}, A.
Caccianiga\altaffilmark{1}, P. Severgnini\altaffilmark{1} and R. Della Ceca\altaffilmark{1}}

 \affil{$^1$ INAF - Osservatorio Astronomico di Brera, via Brera 28, I-20121 Milan, Italy
(braito@brera.mi.astro.it,  tommaso@brera.mi.astro.it, caccia@brera.mi.astro.it, paola@brera.mi.astro.it, rdc@brera.mi.astro.it)} \shorttitle{The first optical  validation of an XLEO} \shortauthors{Braito et
al.}

\begin{abstract}

We present the first optical validation of an X-ray Line-Emitting Object (XLEO) discovered in the XMM-{\it Newton}
observation of the Chandra Deep Field South. The object is an AGN at z$_{x} \sim 0.66$.  An
optical spectrum of the source indeed confirms and refines (z$_{o} = 0.665$)  the redshift
obtained from the X-ray line. The X-ray and optical properties of the source are presented and
discussed.

\end{abstract}

\keywords{Galaxies: Active $-$ Galaxies: Nuclei $-$   X-Rays: Surveys $-$ X-Rays: Galaxies}

\section{Introduction}

We have recently developed FLEX (Finder of Line Emitting X-ray sources),
a new algorithm to search for X-ray Line Emitting Objects (hereafter
XLEOs) in XMM-{\it Newton} observations (Maccacaro et al., 2004). The 
detection  method is  based on a raster scan that is extended  to the
third dimension (energy) that characterizes the EPIC-pn images. In
simple words it is a sliding ``cube'' search algorithm.  Basically, the
counts of the EPIC-pn  data-cube are grouped both in spatial bins and in
energy bins and a search for a significant excess is performed along the
energy axis. A full description of FLEX will be presented in Braito et
al., (in preparation). The first three sources found with this
technique, from the analysis of 13 XMM-{\it Newton} observations, have
been reported and discussed in Maccacaro et al., (2004). The FLEX
algorithm is still being tested in order to optimize, among other
things, its various critical parameters as well as to establish its
efficiency. Over the last few months FLEX was run on a second set of
XMM-{\it Newton} EPIC-pn observations with different combinations of
spatial ($40^{\prime\prime}\times 40^{\prime\prime}$ and
$50^{\prime\prime}\times 50^{\prime\prime}$) and energy sizes (200 and
300 eV) for the three-dimensional detection cell. A substantial number
of XLEOs have been found and this sample will be reported elsewhere.
Since the detection of the X-ray line rests on very few photons (e.g. a
dozen or even less, when 1 or 2 are expected from the continuum) we are faced with the
problem of finding strong independent support to the reality of the
XLEOs detected, especially in this early phase of the project. One of
the strong point of the detection of an XLEO is its redshift
determination directly from the X-ray data, from the assumption that the
line seen is Fe $K\alpha$ at 6.4 keV. The identification of an XLEO with
an object at the same redshift indicated by the iron line, but optically
determined, should thus be considered a strong validation of the reality
of the XLEO. To this end we have   successfully  proposed VLT observations of the
optical counterparts of some of the XLEOs found so far. While waiting
for the observations to be taken, we have decided to run FLEX on a field
for which deep optical spectroscopic observations are already available
for a large number of objects, i.e. the Chandra Deep Field South (CDFS).  For this field  
 deep XMM-{\it Newton} observations exist. In this paper we
present     the first optical validation of an XLEO  deriving from the
discovery of an object coincident with a spectroscopically identified 
X-ray source, having the same redshift as determined  by the FLEX algorithm.

\section {XMM-{\it Newton} observations of the CDFS}

Eight separate XMM-{\it Newton} observations of the CDFS region are available in the
public
archive\footnote{xmm.vilspa.esa.es/external/xmm\_data\_acc/xsa/index.shtml}.
These data have been cleaned from high  background time intervals and the
resulting  pn exposures are between 24 and 64 ks. The log of the observations is
given in Table 1.

We have run FLEX on the 8 individual EPIC-pn observations
and we have selected only those XLEO candidates lying on
a clearly detected XMM-{\it Newton} source.  XLEO
J0332$-$2744 was detected in the analysis of observation
n. 7 (see Table~1). The discovery scan (a box of
$40''\times 40''\times 200$ eV was used) is shown in
Figure 1. An excess in the count distribution is clearly
visible at $\sim$3.8 keV, where 9 counts are recorded.
The ``noise'' level (background + source continuum) is
estimated at 1.06 counts.   Thus the   probability
that the excess is a random noise fluctuation at the
energy and spatial position where it is seen  is
P$_r=1.8\times10^{-6}$. Taking into account that the
total number of trials is $\lae 2000$, this converts into a
probability of finding such a fluctuation in the
observation analyzed of less than $\sim 4\times 10^{-3}$  (see the
discussion in Maccacaro   et al. 2004). A significant
excess was also detected from the same observation in a
scan with a box of $40''\times 40''\times 300$ eV. In
this case 12 counts are  recorded at $\sim$3.8 keV, to be
compared with an expectation of 1.95 (P$_r = 1.0 \times
10^{-6}$).

FLEX did not detect XLEO J0332$-$2744 in any of the other 7
observations. In two cases (n. 1 and n. 3), this is due to the fact
that the source falls in a gap between two CCD chips. In observation n.
8 there is an excess in the count distribution at $\sim$3.7 keV,
however its significance ($3.4\times10^{-4}$) is below the FLEX
threshold for flagging. No excess is seen in the other four
observations.  Prompted by the conviction that the line is indeed
real (an excess of counts at $\sim$ 3.8 keV is seen also in the sum of
the corresponding  MOS observations) we have further analyzed the other
datasets. We determined that the putative emission line which  is
present in the two chronologically consecutive EPIC-pn observations n.
7 \& 8, when the 2$-$10 keV source flux continuum is at a minimum
($<F_x> = 2.3\pm 1.1 \times 10^{-15} $ ergs cm$^{-2}$ s$^{-1} \ $),  is not detectable, if at
the same intensity,  in the other EPIC-pn observations characterized by
a  higher continuum ($<F_x> = 4.8\pm 1.4 \times 10^{-15}$ ergs cm$^{-2}$ s$^{-1} \ $).  
To  summarize, FLEX  detects the line in observation n. 7 as well as in the sum
of  7 \& 8 (15 counts vs 3.3    expected)
but does not detect it in the sum of all the XMM-{\it Newton}
observations. Given the presence of an excess also in the independent
MOS data we conlcude that 
the source is probably variable and that the line is real and best visible
(maximum contrast) when the continuum is low. Data points are however too scanty,
and the source too faint, to allow any investigation of a possible
correlation, or time delay, between continuum and line flux variations.

If the line detected  is interpreted as the Fe  $K\alpha$ emission line at 6.4 keV, then
the redshift of the source can be derived directly from the X-ray data. A proper
spectral analysis (see section 3.1) indicates a best fit energy position for the
line at 3.85$^{+0.45}_{-0.11}$ keV. We thus derive a redshift $z_x =
0.66^{+0.05}_{-0.17}$.

The FLEX algorithm is not providing an accurate X-ray position for the sources found. However, since
XLEO J0332$-$2744 is found at the position of a catalogued Chandra source (Giacconi et al. 2002; 
Szokoly et al. 2004), a very accurate localization of the X-ray source is available ($R.A.=$03:32:36.7;
$Dec=-$27:44:7.1, J2000.0). The XMM-{\it Newton}  position, as derived from the SAS, is consistent,
within 1$^{\prime\prime}$, with the Chandra position. An optical image  of the region of XLEO J0332$-$2744 is
avaiable at: www.mpe.mpg.de/CDFS/data/33.html. The  object  within the  circle  has been already
proposed as the  counterpart of the X-ray source (Giacconi et al., 2002). The optical spectrum 
(from  http://cencosw.oamp.fr/VVDS/CDFS.html) is reproduced  in Figure 2; it shows two prominent
emission lines identified as  [OII]3727\AA~and [OIII]5007\AA \ from which a redshift $z_o = 0.665$ is
derived (Szokoly et al., 2004). This is  in excellent agreement with the X-ray determination, and thus
validates our XLEO detection. 

We have also inspected the 1Ms Chandra ACIS-I observation. The resulting 2-10 
keV
flux is $\sim 2.6$ times higher than the XMM-Newton flux, further implying 
source
variability. Variability (within the Chandra observation) is also 
found (Tozzi, private communication). No evidence for a significant excess is 
found in the
1Ms Chandra observation at $\sim 3.8-3.9$ keV, in agreement with the 
result of Tozzi et
al. 2005 (A\&A, submitted). As we did for the XMM-Newton 
observations
characterized by a high flux, we tested the visibility of a 
line of the same
intensity of that seen in the XMM-Newton observations n. 
7 \& 8 in the ``high
state'' Chandra observation. We found that it would have 
been undetectable. It
must be said that an excess of counts is seen in the 
Chandra data at a sligthly
higher energy: 4.1 keV. Whether this is a real 
emission line and, if so, whether
the 200 eV difference is due to physical 
rather then instrumental reasons is not
clear to us and would require a 
detailed and accurate analysis of the Chandra
data that goes beyond the 
scope of this paper. Given the concordance between the
XMM-Newton EPIC-pn, 
MOS and optical data, the XMM line detection at 3.85 keV
rests on firm 
ground.

\section{The nature of XLEO J0332$-$2744}

\subsection{X-ray Spectral  analysis}
  
To investigate the spectral properties of XLEO J0332$-$2744 we have considered all the 8 public 
XMM-{\it Newton} pointings on this field in the XMM-{\it Newton} archive\footnote{In 2 XMM-{\it Newton}
pointings (n. 1 and n. 3) only MOS1 and MOS2 were used since in the pn the source falls in a gap between two
CCDs.}.   Source counts from each individual exposure and from each instrument (pn, MOS1, and MOS2) 
were extracted using a circular region of radius 13$^{\prime\prime}$ to 15$^{\prime\prime}$, while 
background counts were derived from a nearby  source-free region having an extraction radius 2--3 times
larger. Response matrices at the position of the source have been created using the  SAS tasks {\it
arfgen} and {\it rmfgen}. Summed source and background spectra, as well as summed response matrices and
effective areas, were also created using standard FTOOLS tasks.   

The resulting total exposure times are 277, 416 and 419 ksec for pn, MOS1 and
MOS2 respectively.  The source spectra have been rebinned to have at least
15-20 counts per bin and, to improve statistic, MOS1 and MOS2 data were
combined  together. All  models discussed here have been filtered for  the
Galactic absorbing column density along the line of sight ($N_\mathrm{H} = 8.9
\times 10^{19}$ cm$^{-2}$; Dickey \& Lockman, 1990). Unless otherwise  stated,
errors are given at the 90\% confidence level for one interesting 
parameter ($\Delta\chi^{2} = 2.71$). Finally pn and MOS data were fitted
simultaneously keeping their relative normalizations free to vary.

A single absorbed power law model is a good description of the  combined  data sets; this  simple
model, acceptable from a statistical point of view ($\chi^{2}_{\nu} = 0.82$), gives parameters
which are typical of a relatively unabsorbed AGN ($\Gamma=1.71\pm 0.20$, $N_{\rm H}\sim
1.8\times10^{21}$ cm$^{-2}$),  in agreement with the Chandra X-ray classification  (AGN1, Giacconi
et al. 2002).  However, since an excess of counts  is  present at $\sim 3.8$ keV,  both in the  pn
and MOS data, we have fitted    this ``line-like" excess  adding a narrow gaussian line.   The
addition of the line is not strictly required. However since we have external evidence for the
presence of a line, its inclusion, that leads to a marginal improvement of the fit (see Table 2),
provides a better description of the data. \\

The best fit spectral parameters are reported in Table 2; the best fit  energy position of the line is  3.85 keV, in
very good agreement  with the energy position derived by FLEX, while the line equivalent width ($EW_\mathrm{obs}$) is
$\sim 500$ eV, corresponding to  $EW_\mathrm{rest}$ of $\sim 800$ eV.\\ The spectrum shown in Figure 3    is the
result of the analysis of all  the XMM-{\it Newton}      observations  (upper panel),  and of observations n. 7 plus
8, where    the line has a  higher contrast with respect to the continuum (lower panel). As expected, the line EW 
(see Table 2)  in this latter  case is   larger than in the former, characterized by a   higher continuum.\\

Given the large uncertainty on the equivalent width of this line  it is
difficult to understand if it is  consistent or not with  the measured
intrinsic $N_\mathrm{H}$. If produced by transmission, an $EW_\mathrm{rest}$ of  $\sim  800$
eV implies an absorption column density higher  than $10^{23}$ cm$^{-2}$ 
\citep{turner97,Leahy93}, in contrast with the low absorption measured  from
spectral  fitting. We checked  the possible  presence of an ``extra'' X-ray
absorption  using two different models: adding  a second  absorbed power law
component  or    fitting a partial-covering model.  Both  tests  show
that a  column  density of the order of $10^{23}$ cm$^{-2}$ is  not required 
by the present data.   On the other hand we cannot exclude from the X-ray data
alone that the primary  AGN emission  is  even more absorbed ($\sim 10^{25}$\
cm$^{-2}$) and that its signature falls outside the XMM-{\it Newton} bandpass.

\subsection{Optical data}

The optical spectrum, from the VVDS catalog, shows a continuum emission dominated by the  host-galaxy,
plus two relatively strong  ($EW_\mathrm{rest}\sim$20-30\AA) emission lines ([OII]3727\AA ~~and
[OIII]5007\AA).   The rest--frame wavelength range sampled is 3000\AA--5400\AA.  Given the relative
importance of the host galaxy emission, the properties of the nuclear emission are difficult to
assess.  For this reason, we have adopted a   two-components model, characterized by an AGN emission
plus a galaxy  template, and we attempted to reproduce the observed spectrum, following the approach 
discussed in  Severgnini et al. (2003).  In particular, we have used  an AGN template from Francis et
al. (1991) and  Elvis et al. (1994) and a galaxy template taken from Bruzual \& Charlot (2003). The AGN
narrow emission line ratios are taken from Krolik (1999). In the AGN template  the continuum and the
broad emission lines have been absorbed on the basis of the $N_\mathrm{H}$ value, assuming a Galactic
standard value of E$_{B-V}/N_\mathrm{H}$=1.7$\times$10$^{-22}$ mag cm$^{-2}$ (Bohlin et al. 1978). Both
the normalizations and the value of $N_\mathrm{H}$ are free parameters.

The optical observed spectrum can be well reproduced with an intrinsic column density higher than
3$\times$10$^{21}$ cm$^{-2}$ embedded in a host galaxy. We find that the observed  [OII]3727\AA~ must
be mostly produced by the host galaxy while the [OIII]5007\AA\ probably comes from the AGN.  This model
reproduces also the R$-$K$\sim$4 color measured by Szokoly et al. (2004). The lower limit obtained for
the $N_{\rm H}$ value is  in agreement with the X--ray spectral analysis result. The optical--to--X--ray
spectral index estimated assuming an obscuration of a few  10$^{21}$ cm$^{-2}$
($\alpha_{ox}$\footnote{The $\alpha_{ox}$ is defined  in the rest frame as follows:
$\alpha_{ox}=-$log(f$_{2500\rm \AA}$/f$_{2\rm keV}$)/log($\nu_{2500\rm \AA}$/$ \nu_{2\rm
keV})$.}$\sim$1.4) is similar to the typical value expected for a high--luminosity Seyfert
($\alpha_{ox}\sim1.5$). This model does not put any upper limit on the amount of obscuration. 
However, assuming a strong X--ray obscuration of $N_\mathrm{H}=$10$^{24}$ cm$^{-2}$ 
or more, the intrinsic 2 keV flux would be at least a factor of 100 higher and the
$\alpha_{ox}$ spectral index would be $<$1. This latter value, typical of {\it
low luminosity AGN} (e.g. Ho 1999), is in contrast with the high intrinsic
2--10 keV luminosity  that this source would have (i.e. in the  ``Compton thick''  hypothesis  the intrinsic 2--10 keV
luminosity  would be $>$10$^{44}$ erg s$^{-1}$).

An additional constraint on the total amount of obscuration can be derived from
the thickness parameter (T=F$_{2-10 keV}$/F$_{[OIII]}$) discussed in Bassani et al. (1999).
This ratio, combined with the EW of the Fe line, 
is used as an indicator for the presence of strong obscuration 
in local AGN.  In order to apply this diagnostic to our source, we have estimated
the rest--frame 2--10 keV and [OIII] fluxes. We obtain T$\sim$70 which 
excludes the presence of a highly obscured (a ``Compton thick'') AGN which is characterized 
by T$<$1 (see Bassani et al. 1999).

In summary, the optical properties of the source are all consistent with a
scenario in which an AGN with an $N_\mathrm{H}$ value between a few  10$^{21}$ cm$^{-2}$ to a few 10$^{23}$ cm$^{-2}$ is present. On the basis of our
model, this scenario could be firmly confirmed  and refined by observing the spectral region were the H$\alpha$ line is 
expected (i.e. around 1.1 $\mu$). In the case of low absorption, a strong and broad
H$\alpha$ line should be observed, while no broad component should be present
in the case of strong obscuration.

\section{Discussion and conclusions}

We have presented the first optical validation of an XLEO, namely an independent optical
redshift determination that confirms the redshift derived from the X-ray data alone. XLEO
J0332$-$2744 was discovered in the XMM-{\it Newton} observation of the Chandra Deep Field
South. It is coincident with a previously reported Chandra and XMM-{\it Newton} source,
identified with an AGN at moderate redshift (z = 0.665). From the X-ray and optical spectra
the object seems to be only moderately absorbed ($N_{\rm H}\sim 3\times 10^{21}$ cm$^{-2}$)
unlike the first 3 XLEO discovered that were close to the ``Compton  thick'' regime, being
characterized by $N_\mathrm{H} \gae 10^{24}$ cm$^{-2}$ (Maccacaro et al. 2004). Deeper X-ray observations are
difficult to conceive given the faintness of the source. Observations at other wavelengths
are however feasible and have the potential of providing crucial informations for a better
understanding of the nature of this source. In particular infrared spectroscopy in the
region around $\sim 1.1\mu m$, where H$_\alpha$ in emission is expected, will allow to
better constrain the amount of intrinsic obscuration.\\ It is worth stressing that, for sources with limited
statistics,   the standard spectral   description used  for X-ray data  (i.e. through  binning  of
the original data)   does not always allow to recognize the presence of a  line; indeed the  inevitable large energy binning 
adopted   dilutes the  source counts   at the energy of the  emission line.  The results presented here confirm
the power of FLEX, the algorithm we have specifically developed to search for X-Ray Line
Emitting Objects.

\acknowledgments

\acknowledgments We would like to thank the anonymous referee for his/her useful comments that have contributed improving this
Letter.  We also thank C. Vignali  and A. Wolter for    suggestions     on   the {\it Chandra}  data analysis,  P. Tozzi for useful
discussion and for communicating results prior to publication  and   our system manager C. Bernasconi for her excellent and timely 
support to our computational needs.    PS acknowledges a research fellowship from the Istituto Nazionale di  Astrofisica (INAF). The
XLEO project at the Brera Astronomical Observatory is supported by a MIUR COFIN grant.

\clearpage

\begin{figure}
\epsscale{.60}
\plotone{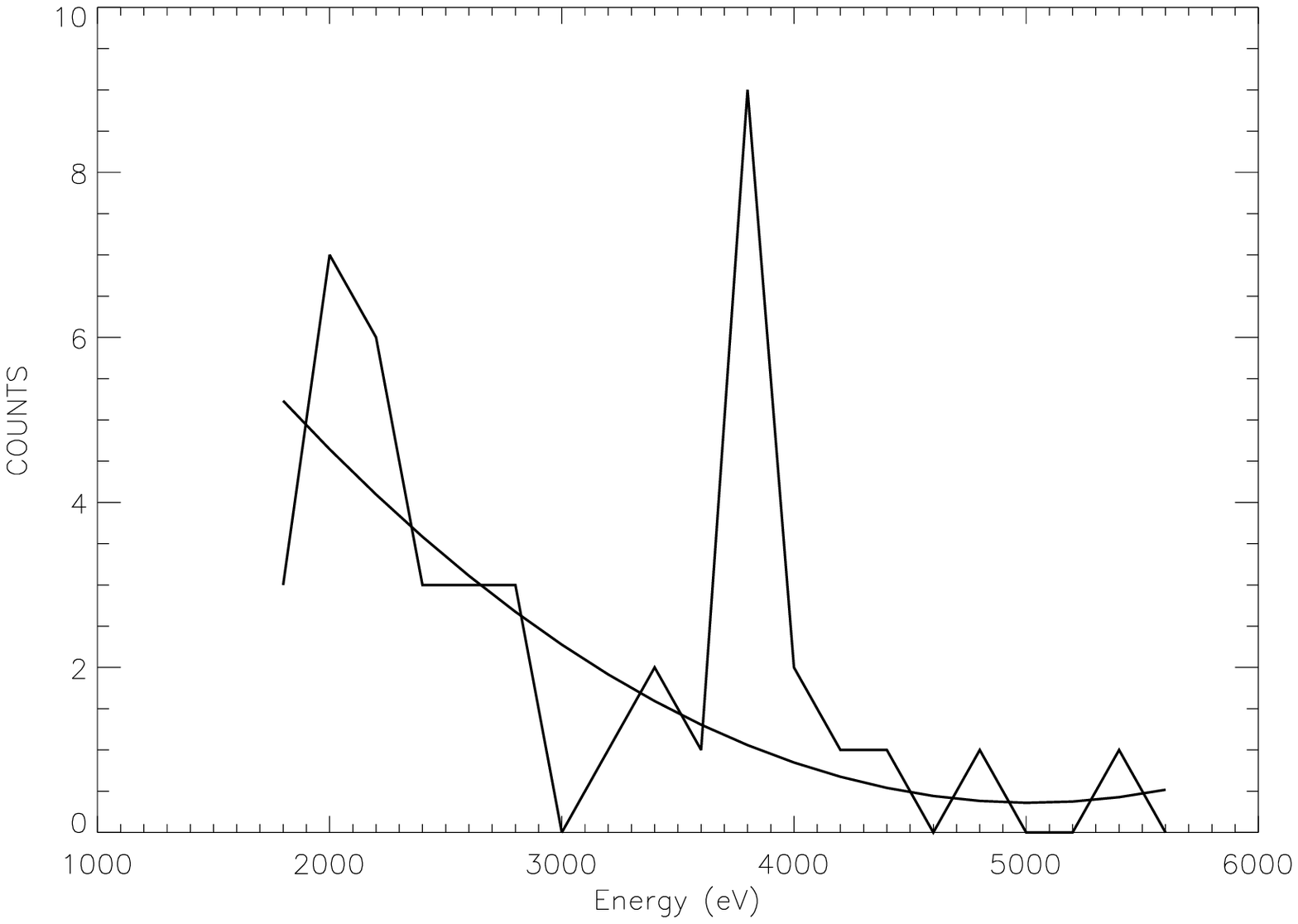}
\caption{Counts distribution from the detection scan. The estimated
``background'' level is indicated.}
\label{fig1}
\end{figure}
 
\clearpage

\begin{figure}
\epsscale{.60}
\plotone{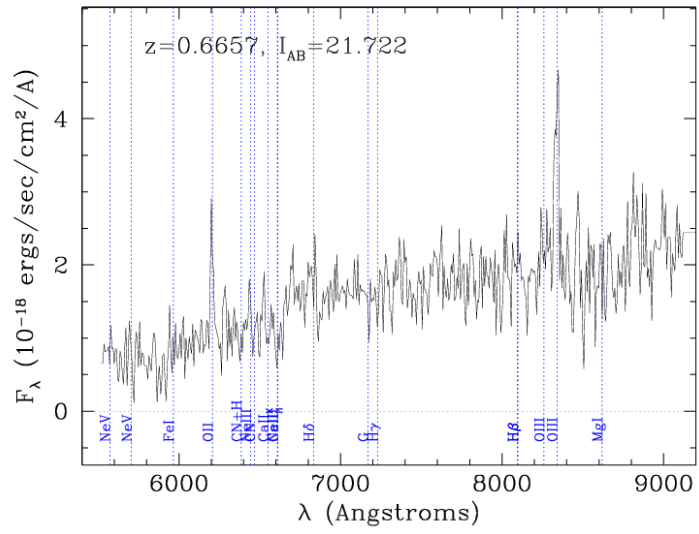}
\caption{Optical spectrum of XLEO J0332$-$2744 retrived from the VVDS catalogue at: http://cencosw.oamp.fr/VVDS/CDFS.html}
\label{fig2}
\end{figure}

\clearpage

\begin{figure} 
\includegraphics[angle=-90,scale=.50]{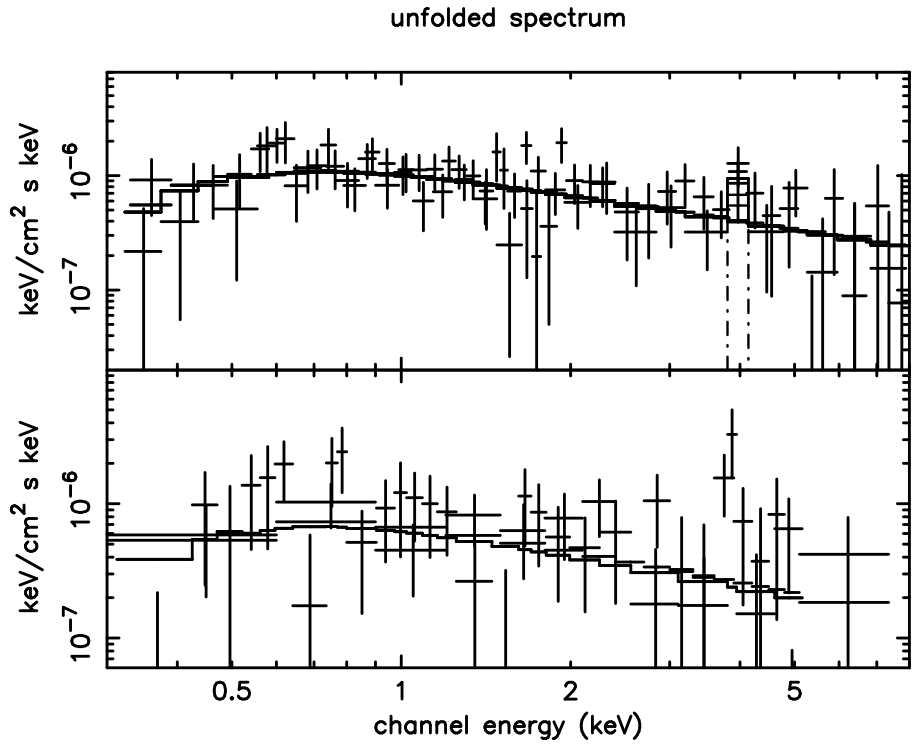} 
\caption{Data points and best fit model  for XLEO J0332$-$2744.  Upper panel: unfolded   pn  and MOS spectrum from 
 all the XMM-{\it Newton} exposures.  Lower panel: same as above   but  for exposures n. 7 plus  8 only. In the
interest of clarity  the gaussian line component of the model has not been plotted.} 

\label{fig3} \end{figure}

\clearpage
 
\begin{deluxetable}{lllll}
\tablecolumns{6}
\tablewidth{0pc}
\tablecaption{Summary of XMM-{\it Newton} observations of the CDFS}
\tablehead{
\colhead{N}    & \colhead{Obsid}    & \colhead{Exposure (s)}    & \colhead{Obs. Date}    & \colhead{Notes}     \\
\colhead{}    & \colhead{ }    & \colhead{pn MOS1, MOS2}    & \colhead{ }        & \colhead{ }    }
\startdata

1& 0108060401& 24686, 30508, 30520  &  2001-07-27  & in pn gap    \\
2& 0108060501& 33592, 45485, 45584  &  2001-07-28  & line not detected     \\
3& 0108060601& 39110, 49004, 49703  &  2002-01-13  & in pn gap    \\
4& 0108060701& 63746, 76808, 77042  &  2002-01-14 & line not detected	  \\
5& 0108061801& 42824, 50480, 52117  &  2002-01-16 & line not detected	  \\
6& 0108061901& 34868, 43621, 43825  &  2002-01-17 & line not detected		      \\
7& 0108062101& 39152, 43601, 43752  &  2002-01-20 & XLEO detected in pn \\
8& 0108062301& 63986, 76608, 76433  &  2002-01-23 & line present but not detected   \\
\enddata
\end{deluxetable}

\clearpage

\begin{deluxetable}{lcc ccc ccl}
\tablecolumns{9}
\tablewidth{0pc}
\tablecaption{Results from spectral fitting of XLEO J0332$-$2744.}
\tablehead{
\colhead{MODEL}    & \colhead{$\Gamma$}    & \colhead{$N^{a}_{\rm H}$}    & \colhead{E$_c$}    & \colhead{EW$^b$}   &
  \colhead{Flux$_{2-10}$}& \colhead{L$_{2-10}$} & \colhead{$\chi^{2}/dof$ }& \colhead{INST } \\&
   &     \colhead{10$^{21}$cm$^{-2}$}    & \colhead{keV }    & \colhead{keV}   &\colhead{10$^{-15}$cgs}& \colhead{10$^{43}$cgs} &}
\startdata
PL 	       &1.71$^{+0.18}_{-0.20}$ 	&  1.8$^{+1.6}_{-1.1}$&/&/&  4.6$^d$&0.7&60.6/74& XMM all\\

PL+Gauss.       &1.78$^{+0.34}_{-0.25}$ &  1.8$^{+2.0 }_{-1.1 }$&3.9$^{+0.4 }_{-0.1}$&$0.5^{+0.5}_{-0.4}$&  4.5$^e$&0.8&57.6/72&  XMM
all\\
PL+Gauss.       &1.8$^c$ &  1.8$^{+2.4 }_{-1.8 }$&3.9$^{+0.1}_{-0.1}$&$1.9^{+1.4}_{-1.3}$ &  3.2$^e$&0.6&40.2/41& XMM 7$\&$8\\

\enddata
\tablenotetext{a}{Rest frame}
\tablenotetext{b}{Observed frame}
\tablenotetext{c}{This parameter has been  frozen due to the low  statistics. }
\tablenotetext{d}{This flux refers to the continuum of the source. }
\tablenotetext{e}{This flux includes also the contribution of the Fe line emission. }

\end{deluxetable}
 
\end{document}